\documentclass{kluwer}    

\usepackage{psfig}

\newdisplay{guess}{Conjecture}

\begin{document}                                                                                   
\begin{article}
\begin{opening}         
\title{Black Hole Spin in AGN and GBHCs}
\author{Christopher S. \surname{Reynolds}$^1$, Laura W. \surname{Brenneman}$^1$, David \surname{Garofalo}$^2$}  
\runningauthor{C.~S.~Reynolds, L.W.Brenneman, D.Garofalo}
\runningtitle{Black Hole Spin}
\institute{$^1$Dept. of Astronomy, University of Maryland, College Park, MD20742$^1$\\Dept. of Physics, University of Maryland, College Park, MD20742}
\date{}

\begin{abstract}
  We discuss constraints on black hole spin and spin-related
  astrophysics as derived from X-ray spectroscopy.  After a brief
  discussion about the robustness with which X-ray spectroscopy can be
  used to probe strong gravity, we summarize how these techniques can
  constrain black hole spin.  In particular, we highlight {\it
    XMM-Newton} studies of the Seyfert galaxy MCG--6-30-15 and the
  stellar-mass black hole GX~339--4.  The broad X-ray iron line
  profile, together with reasonable and general astrophysical
  assumptions, allow a non-rotating black hole to be rejected in both
  of these sources.  If we make the stronger assertion of no emission
  from within the innermost stable circular orbit, the MCG--6-30-15
  data constrain the dimensionless spin parameter to be $a>0.93$.
  Furthermore, these {\it XMM-Newton} data are already providing
  evidence for exotic spin-related astrophysics in the central regions
  of this object.  We conclude with a discussion of the impact that
  {\it Constellation-X} will have on the study of strong gravity and
  black hole spin.
\end{abstract}

\end{opening}           

\section{Introduction}

With evidence for the existence of black holes and the dynamical
measurement of their mass becoming almost pass\'e, an increasing focus
is being placed on detecting the effects of black hole spin.  Spin
truly is a creature of the relativistic Universe, and the
observational investigation of spin puts us one step closer to being
able to genuinely test strong-field General Relativity (GR).  Even if
GR passes all of these tests (which, of course, would be the most
``boring'' possibility), black hole spin gives us crucial insight into
how black holes of all masses are born, and may well be an important
ingredient in powering some of the most energetic sources in the
Universe.

At the current time, the best evidence for the effects of black hole
spin come from X-ray observations, both timing and spectroscopy.
X-ray variability studies, particularly investigations of
quasi-periodic oscillations (QPOs) have produced tantalizing hints
that we might be witnessing the effects of black hole spin (Stella,
Vietri \& Morsink 1999; Strohmayer 2001).  However, the lack of any
agreed upon theoretical framework for the high-frequency QPOs prevents
us from drawing robust conclusions at this time.  For this reason,
the most compelling studies of black hole spin have originated from
X-ray spectroscopy.

In this contribution, we describe constraints on black hole spin from
X-ray spectroscopy.  We will align our discussion around three
questions; ``Have we seen the effects of strong gravity at all?'',
``Have we seen the effects of black hole spin?'', and ``Can we probe
the exotic astrophysics associated with spinning black holes?''  For
the impatient reader, the answers to these questions are ``Yes!'',
``Very probably'', and ``We're maybe just starting to...''.  We
conclude by discussing future prospects for probing black hole spin
and testing strong-field GR with both X-rays and gravitational waves.

\section{Strong gravitational effects in X-ray spectra}

The principal spectroscopic tool used to date to study strong gravity
is the characterization of the broad iron-K$\alpha$ fluorescent
emission line (see reviews by Fabian et al.  2000 and Reynolds \&
Nowak 2003).  The essential physics underlying this phenomenon is
straightforward.  Moderate-to-high luminosity black hole systems
accrete via a radiatively-efficient disk.  Even in the region close to
the black hole, such a disk will (apart from a hot and tenuous X-ray
emitting corona) remain optically-thick, geometrically-thin, almost
Keplerian, and rather cold ($T<10^7$\,K).  X-ray irradiation of the
surface layers of the disk by the corona will excite observable
fluorescence lines, with iron-K$\alpha$ being most prominent due to
the combination of its astrophysical abundance and fluorescent yield.
This emission line is then subject to extreme broadening and skewing
due to the both the normal and transverse Doppler effect (associated
with the orbital velocity of the disk) as well as the gravitational
redshift of the black hole (see Fig.~1).

So, have we seen these effects in the X-ray spectra of real accreting
black holes?  Broad emission features that can be modeled as iron
emission lines from the central regions of a Keplerian accretion disk
are present in the {\it XMM-Newton} data for over half of the
moderate-to-high luminosity Seyfert galaxies, as well as many Galactic
Black Hole Candidates (GBHCs) in their intermediate and high state.  Given the
breadth of these features, it is valid to ask whether continuum
curvature and/or unmodeled complex absorption might be mimicking a
broad emission line.  In some cases, detailed scrutiny of high
signal-to-noise {\it XMM-Newton} data allows one to reject these
alternatives, further validating the relativistic line interpretation
(e.g. MCG--6-30-15 [Vaughan \& Fabian 2004, Reynolds et al. 2004, also
see Fabian et al.  1995], GX~339--4; Miller et al.  2004).  In other
cases, absorption by large columns of photoionized material appears to
be important.  In the case of NGC~4151, for example, much of the broad
iron line reported by Wang et al.  [1999] was probably an artifact of
not modeling the complex absorber later identified by Schurch \&
Warwick (2002) and Schurch et al. (2003).  In many other cases, the
role that complex absorption has on the presence of a broad iron line
remains uncertain.  It is important to stress, however, that the role
of photoionized absorption in masking or mimicking broad iron lines is
{\bf knowable}, and will be elucidated by the high-resolution and high
count-rate spectra that {\it Astro-E2} will be obtaining on a regular
basis starting in early 2005.

To summarize this section, there are robust examples of broad iron
emission lines that are giving us a clean probe of the strong gravity
region around both stellar and supermassive black holes.  However,
while broad iron lines are not rare, the precise fraction of objects
in the various classes of accreting black holes that display these
features is still uncertain.  In addition to obtaining new
high-resolution and high count-rate spectra with {\it Astro-E2},
significant progress is possible in this field via a large and
unbiased survey of current {\it XMM-Newton} data.

\section{Constraining black hole spin with X-ray spectroscopy}

Having established that at least some accreting black holes display
broad iron lines that cleanly probe the strong gravity region, we now
ask whether we can constrain the black hole spin using these features.
To sharpen the discussion, we will address whether one can rule out a
Schwarzschild metric (i.e., non-spinning black hole) for any given
black hole system.

Even with {\it XMM-Newton}, we cannot probe the iron line on the
dynamical timescale of the very centralmost regions of the accretion
disk where spin effects are dominant.  We are driven to study line
profiles that have been time-averaged over several dynamical
timescales --- hence, our primary information on black hole spin at
the current time will originate from the breadth and redshift of these
time-averaged line profiles.

There is a common misconception that rapidly spinning black holes
invariably produce broader and more highly redshifted emission lines
than slowly spinning black holes.  This stems from the fact that the
innermost stable circular orbit (ISCO; $6GM/c^2$ for a non-rotating
black hole) for a prograde accretion disk pulls-in towards the horizon
as the spin-parameter of the black hole is increased. Hence, the line
broadening will increase with black hole spin {\it if} the line
emission is always truncated at the ISCO.  But it is important to
realize that we can produce arbitrarily redshifted and broadened
emission lines from around even a non-rotating black hole {\it if
  nature had the freedom to produce line emission from any radius
  beyond the horizon} (Reynolds \& Begelman 1997).  This discouraging
fact has led some authors to conclude that current iron line profiles
contain essentially no information on the black hole spin (Dovciak,
Karas \& Yaqoob 2004).

This would be an overly bleak assessment of our ability to
constrain black hole spin.  Even the application of some rather weak
(i.e., general) astrophysical constraints can impose an inner limit on
the radii at which spectral features can be produced.  In order to
produce any significant iron emission line from the region within the
ISCO (which we shall refer to as the plunging region), the disk in
this region must be optically-thick, not too highly ionized (i.e., a
significant fraction of the iron cannot be fully ionized), and
illuminated by the hard X-ray continuum.  While much work remains to
be done on the physical state of matter in the plunging region, it is
challenging to construct a model for a non-rotating black hole in
which there are appreciable spectral features produced by matter
inside of $4-5GM/c^2$ (Reynolds \& Begelman 1997).  If we require an
emitting radius less than this when fitting a non-rotating black hole
model to a particular dataset, we can claim to have found good
evidence for a spinning black hole.

This is exactly the situation we find when attempting to fit the {\it
  XMM-Newton} data for the Seyfert-1 galaxy MCG--6-30-15.  The
June-2000 observation of this source (reported by Wilms et al. [2000]
and Reynolds et al. [2004]) caught it in its enigmatic ``Deep Minimum
State'' first identified with {\it ASCA} data by Iwasawa et al.  (1996)
during which the iron line is known to be particularly broadened and
redshifted.  Fitting the Reynolds \& Begelman (1997) Schwarzschild
iron line model which includes emission from within the plunging
region results in essentially all of the emission being placed at
$3GM/c^2$.  It is extremely hard to understand how this could be a
physical result --- the relativistic inflow at this location demands
(through mass continuity) that the density be low and, hence, that
this material be completely photoionized if it were to experience the
irradiation suggested by this fit.  Thus, the extreme parameters
derived from a fit to a Schwarzschild-based model leads to the conclusion
that we are seeing the effects of black hole spin.

Given an observation of a very broad iron line such as that detected
in the Seyfert-1 galaxy MCG--6-30-15 or the GBHC GX~339--4 (Miller et
al. 2004), we can place constraints on the black hole spin given
certain astrophysical assumptions.  The systematic exploration of
these constraints has only just begun and is still a work in progress.
To facilitate this work, we have constructed a new iron line profile
code {\tt kerr} (that employs the Kerr metric ray-tracing code of
Speith, Riffert \& Ruder 1995) which treats the black hole spin as a
free parameter.  This code also takes advantage of modern computing
speeds and performs the necessary calculations in real time as the
spectrum is being fit with {\sc xspec}.  The user may therefore tune
the spectral resolution and numerical accuracy of the model to suit
the data at hand, a feature that is not available in the tabular
models such as {\tt laor} that have been extensively employed to date.

\begin{figure}
\centerline{
\psfig{figure=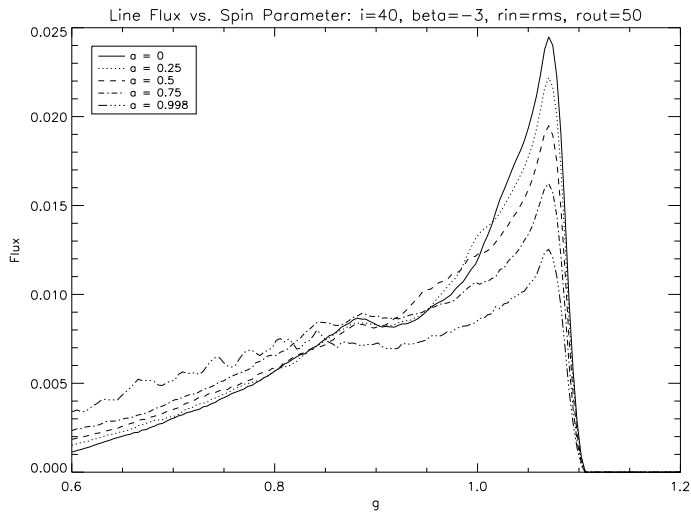,width=0.8\textwidth,angle=90}
}
\caption{Theoretical lines profiles from a Keplerian accretion disk 
  around a Kerr black hole.  We assume an observed inclination angle
  of $i=40^\circ$ and a line emissivity that falls of as $r^{-3}$
  between the ISCO and $r_{\rm out}=50GM/c^2$.}
\end{figure}

Preliminary fitting of {\tt kerr} to the highest signal-to-noise data
for MCG--6-30-15 (from the June-2001 observation) demonstrates that
the black hole must possess a dimensionless spin parameter of $a>0.93$
(Brenneman \& Reynolds, in prep) {\it if we impose the condition that
  no spectral features are produced from within the ISCO.}  See Fig.~1
for examples of line profiles calculated under this assumption.
Current work is focused on obtaining spin constraints once that
assumption is relaxed.  Note that these fits assume a broken power-law
form for the line emissivity as a function of radius.  Hence, our
limit of $a>0.93$ is a stronger constraint than the $a>0.94$ deduced
by Dabrowski et al. (1997) from {\it ASCA} data who assumed the line
emissivity tracks the radial dissipation profile of a ``standard''
(Novikov \& Thorne 1974; Page \& Thorne 1974) accretion disk.  As will
be discussed below, the {\it XMM-Newton} data are of sufficient
quality to actually falsify the Dabrowski et al. assumption.

\section{The exotic astrophysics of spinning black holes}

Rapidly-spinning black holes are undoubtedly amongst the most exotic
objects in the current-day universe.  In this section, we focus on one
particular facet of their behaviour --- the magnetic interactions
between the spinning black hole and surrounding matter including the
accretion disk.  We argue that {\it XMM-Newton} data are already
hinting at evidence for the magnetic extraction of spin energy from
the black hole in MCG--6-30-15.

Analytic (Krolik 1999, Gammie 1999) and numerical (Hawley \& Krolik
2001, Reynolds \& Armitage 2002) studies have shown that magnetic
forces can couple material within the plunging region to the body of
the accretion disk, thereby extracting energy and angular momentum
from that region.  In an extreme limit, a Penrose process\footnote{We
  note that Williams [2003] has also argued for the importance of a
  non-magnetic, particle-particle and particle-photon scattering
  mediated Penrose process.} might be realized in which the innermost
regions of the accretion flow are placed on negative energy orbits by
these magnetic torques (Agol \& Krolik 2000).  Together with any
Blandford-Znajek process (Blandford \& Znajek 1977) that might result
from field lines directly connecting to the (stretched) horizon, these
magnetic torques can in principal extract a black hole's spin energy,
depositing it either in the body of the disk or in the form of an
outflow of mass and/or Poynting flux.  Note that all of this behaviour
is in stark contrast to standard black hole disk models (Shakura \&
Sunyaev 1973, Novikov \& Thorne 1974, Page \& Thorne 1974) in which
material follows conservative orbits once inside the ISCO.

\begin{figure}
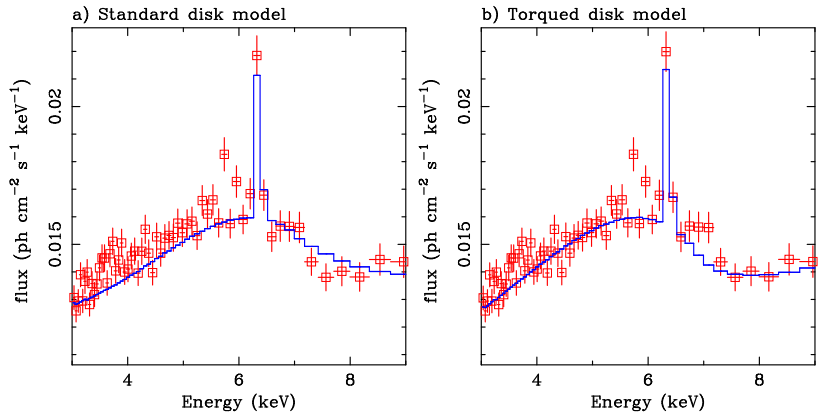

\hbox{
\psfig{figure=f2a.ps,width=0.45\textwidth,angle=270}
\psfig{figure=f2b.ps,width=0.45\textwidth,angle=270}
}
\caption{Broad iron line fit assuming that the line emission tracks the 
  underlying disk dissipation of (a) a standard (Novikov \& Thorne
  1974) accretion disk and, (b) an Agol \& Krolik (2000) torqued
  accretion disk.  Modified from Reynolds et al. (2004).}
\end{figure}

Can we see evidence for any of these processes in the current data?
Again, we return to the Deep Minimum State of MCG--6-30-15 which
displays one of the broadest and most highly redshifted iron lines
known.  This immediately tells us that the X-ray reflection features
are originating from a region that is extremely centrally-concentrated
in the accretion disk.  For the moment, we assume that the primary
continuum X-ray source is located a small distance above the disk
surface (the ``local corona approximation'') and radiates a fixed
fraction of the energy dissipated in the underlying disk.  Then, even
assuming a near-maximal rotating black hole (with $a=0.998$), these
data cannot be adequately described with a model consisting of a
standard Novikov \& Thorne (1974) accretion disk --- the model simply
cannot reproduce the centrally concentrated emission pattern inferred
from these data (Fig.~2a).  One can attempt to rescue the standard
disk model by supposing that a larger portion of the total dissipation
in the disk is channeled into the X-ray emitting corona as one moves
to smaller radii.  However, since 30--50\% of the bolometric power of
MCG--6-30-15 seems to emerge through the X-ray emitting corona, one
cannot decouple it entirely from the dissipation distribution.  In the
most extreme model (which provides an adequate but not the best fit to
the data), {\it all} of the dissipated energy is channeled into the
X-ray emitting corona within the central $5GM/c^2$, while the X-ray
production efficiency is zero beyond that radius.

Our best-fitting model consists of a strongly torqued accretion disk
in which the extreme central concentration originates from a magnetic
torque by the plunging region or the rotating black hole (Reynolds et
al. 2004; Fig.~2b) ---  the work done by the torque is
dissipated in the main body of the accretion disk and, with some
efficiency, energizes the inner regions of the X-ray emitting corona.
If this is really the correct description of the physics at play, the
data argue that the accretion disk is in an extreme torque-dominated
state, i.e., the disk is predominately shining through the release of
black hole spin energy.

MHD simulations suggest that magnetic connections between the plunging
region and the body of an accretion disk tend to be rather sporadic.
It is then tempting to identify MCG--6-30-15's transition into the
Deep Minimum State as the (temporary) onset of a significant inner
torque.  The fact that the overall luminosity of a disk necessarily
increases when an inner torque is applied (due to the dissipation of
the extra work done by the inner torque), in contrast to X-ray flux
drop observed during the Deep Minimum State, may be a problem for this
model.  However, the enhanced returning radiation associated with the
torque-induced emission will strongly Compton cool the X-ray corona
leading to a steepening of the X-ray continuum and (possibly) a
large-scale condensation-driven collapse of the corona.  Such effects
may be responsible for the X-ray flux decrease (Garofalo \& Reynolds
2004).

It is also possible that the local-corona approximation is not valid.
If the X-ray emitting source is a significant height above the
optically-thick part of the accretion disk, the hard X-ray continuum
photons will be gravitationally focused into the central regions of
the accretion disk (see Andy Fabian's contribution in these
proceedings).  Aspects of this scenario have been explored by many
authors including Martocchia \& Matt (1996), Reynolds \& Begelman
(1997) and Miniutti \& Fabian (2004).  This suggests an alternative
picture in which the Deep Minimum State is produced when the X-ray
source is located at mid/high latitudes very close to the black hole.
The centrally concentrated X-ray reflection results from the
gravitational focusing, and the decrease in the observed continuum
X-ray flux is a natural consequence of the fact that the continuum
photons are focused {\it away} from the observer (Reynolds \& Begelman
1997; Miniutti \& Fabian 2004).  We note that this scenario does not
diminish the need for exotic spin-related astrophysics --- the base of
a spin-driven magnetic jet is an obvious candidate for this elevated
continuum X-ray source.

\section{Conclusion and the future of black hole studies}

Current data are already allowing us to probe black hole physics
within a few gravitational radii of the event horizon, and may well be
giving us the first observational glimpses of physics within the
ergosphere.  But this is just the beginning of X-ray astronomy's
exploration of strong gravity, not the end of the road.  The enormous
throughput of {\it Constellation-X} will allow us to probe detailed
time variability of the iron line.  Dynamical timescale line
variability, an easy goal for {\it Constellation-X}, will allow us to
follow non-axisymmetric structures in the disk as they orbit (Armitage
\& Reynolds 2003; also see Iwasawa, Miniutti \& Fabian [2003] for the
first hint of such structure in {\it XMM-Newton} data).  This gives us
a direct probe of an almost Keplerian orbit close into a black hole.
Furthermore, line variability on the light crossing time will allow us
to probe relativistic reverberation signatures (Reynolds et al. 1999;
Young \& Reynolds 2000), essentially giving us a direct probe of the
null geodesics in the space-time.  Together, these variability
signatures will allow true tests of strong-field GR.

There is no compelling reason to believe that GR fails on the
macroscopic scales probed by either X-ray or gravitational wave
studies of astrophysical black holes.  In the event that GR is
verified, both X-ray and gravitational wave observations will allow
unambiguous measurements of black hole spins.  Gravitational wave
observations with {\it LISA} of a stellar mass black hole spiraling
into a $10^6\,M_\odot$ black hole (a so-called Extreme Mass Ratio
Inspiral; EMRI) will allow precision measurement of the supermassive
black hole's spin as well as tests of the no-hair theorem and the Kerr
metric.  The event rates of such sources is quite uncertain, however,
partially due to the recent evidence for ``anti-hierarchical'' black
hole growth (e.g., Marconi et al.  2004) and its implications for the
number density of $10^6\,M_\odot$ black holes in the cosmic past.
X-ray spectroscopy with {\it Constellation-X} provides a crucial
parallel track of study in which we can obtain measurements of black
hole spin across the whole mass range of astrophysical black holes
(i.e., stellar, intermediate, and supermassive) using spectral
features that are already known to exist.  Only then can the
demographics and astrophysical relevance of black hole spin truly be
gauged.

We thank the conference organizers for a stimulating and highly
enjoyable meeting.  We also thank the National Science Foundation (US)
for support under grant AST0205990.

\end{article}
\end{document}